# Rare event estimation with sequential directional importance sampling (SDIS)

Kai Cheng [a,b], Iason Papaioannou[c], Zhenzhou Lu[a*], Xiaobo Zhang[a], Yanping Wang[a]

[a]School of Aeronautics, Northwestern Polytechnical University, Xi′an,710072,PR China

[b]Department of Mathematics and Computer Science, University of Southern Denmark, Odense, Denmark

[c]Engineering Risk Analysis Group, Technische Universität München. Arcisstraße 21, 80333 München, Germany

## Abstract

In this paper, we propose a sequential directional importance sampling (SDIS) method for rare event estimation. SDIS expresses a small failure probability in terms of a sequence of auxiliary failure probabilities, defined by magnifying the input variability. The first probability in the sequence is estimated with Monte Carlo simulation in Cartesian coordinates, and all the subsequent ones are computed with directional importance sampling in polar coordinates. Samples from the directional importance sampling densities used to estimate the intermediate probabilities are drawn in a sequential manner through a resample-move scheme. The latter is conveniently performed in Cartesian coordinates and directional samples are obtained through a suitable transformation. For the move step, we discuss two Markov Chain Monte Carlo (MCMC) algorithms for application in low and high-dimensional problems. Finally, an adaptive choice of the parameters defining the intermediate failure probabilities is proposed and the resulting coefficient of variation of the failure probability estimate is analyzed. The proposed SDIS method is tested on five examples in various problem settings, which demonstrate that the method outperforms existing sequential sampling reliability methods.

## 1. Introduction

In the field of reliability analysis, the basic task is to estimate the failure probability $P_f$ by solving the following integral

$$P_f = \int_{g(\boldsymbol{x})\le 0} f_{\boldsymbol{X}}(\boldsymbol{x})\, d\boldsymbol{x} = \int_{\mathbb{R}^n} I(g(\boldsymbol{x}) \le 0) f_{\boldsymbol{X}}(\boldsymbol{x})\, d\boldsymbol{x}, \tag{1}$$

*Corresponding author.

E-mail address: zhenzhoulu@nwpu.edu.cn

where $\boldsymbol{X} = (X_1, \dots, X_n)^{\mathrm{T}}$ are the input random variables with joint probability density function (PDF) $f_{\boldsymbol{X}}(\boldsymbol{x})$, $g(\boldsymbol{x})$ is the limit state function (LSF), failure of the system of interest is defined as the event $\{g(\boldsymbol{X}) \leq 0\}$, and $I(\cdot)$ is the indicator function, i.e., $I(true) = 1$ and $I(false) = 0$. In most engineering applications, failure of the system corresponds to a rare event and, hence, estimating the probability of failure of Eq. (1) corresponds to a rare event estimation problem.

The LSF often depends on the outcome of a "black-box" computationally expensive engineering model, which makes evaluation of the reliability integral a nontrivial task. A number of tailored methods have been developed to efficiently evaluate the integral in Eq. (1) known under the umbrella term structural reliability methods [1]. Existing methods in the literature can be categorized into four types: approximate analytical methods [2, 3], moment-based methods [4-8], surrogate-assisted methods [9-13] and numerical simulation methods [14-16]. Approximate analytical methods expand the LSF $g(\boldsymbol{x})$ at the most likely failure point (aka design point) by a low order Taylor expansion to estimate the failure probability; this category includes the First order reliability method (FORM) [2] and Second order reliability analysis method (SORM) [3, 17]. Although these methods are computationally efficient, their accuracy cannot be guaranteed for highly non-linear and high dimensional problems. Moment-based methods compute the first few moments of the LSF by the point estimation method [4-6] or sparse grid integration method [7], and then estimate the failure probability in terms of the moments. The main disadvantage of these methods is that their computational cost increases fast with the input variable dimension. Surrogate-assisted methods aim at constructing a surrogate model of the computationally expensive engineering model using an explicit mathematical expression based on a set of training points. This enables performing the reliability analysis efficiently based on the cheap-to-evaluate surrogate model. Surrogate models are able to alleviate high computational demands but their performance often deteriorates with increase of the problem dimension. Numerical simulation methods include Monte Carlo simulation (MCS) and variance reduction methods such as importance sampling (IS) [18-20] and its enhanced versions [21-24], subset simulation (SuS) [15, 16, 25], thermodynamic integration and parallel tempering (TIPT) [26], line sampling (LS) [27-29], directional sampling (DS) [30, 31], among others. Simulation methods generally provide robust probability estimates for complex problems, and thus have drawn much attention in the reliability analysis community.

In the present paper, we focus on the sequential importance sampling (SIS) method. SIS is also known as bridge importance sampling, and it has been implemented for reliability analysis and design optimization in [21, 22]. Also, the popular SuS method can be regarded as a particular case of SIS [21, 32]. The common idea of these methods is to define a set of intermediate densities known up to a normalizing constant, such that the final density approximates well the theoretically optimal IS density. These densities are sampled sequentially through application of Markov chain Monte Carlo (MCMC) algorithms, resulting in a step-wise exploration of the failure region. The final estimate of the probability of failure is expressed as a function of estimates of the normalizing constants, which are obtained in each sampling step. As a result, a complex reliability analysis problem can be decomposed into a series of simpler problems. Here we present a new SIS method in polar coordinates, which we term sequential directional importance sampling (SDIS) for reliability analysis. In SDIS, the failure probability is expressed as a function of a set of auxiliary probabilities, in which the first one is estimated with MCS in Cartesian coordinates, while the subsequent ones are estimated with directional importance sampling in polar coordinates. The MCS population in the first step is used to exploit the parameter space globally, while MCMC sampling is utilized in the subsequent steps to explore the local failure regions of interest in a sequential manner. The performance of the proposed algorithm is demonstrated with several benchmark examples.

The layout of this paper is as follows. Section 2 reviews the background of MCS, IS, DS and directional IS (DIS). In Section 3, the basic idea and rationale of SDIS is first presented. Then, specific ingredients for its efficient implementation, including a resampling strategy and two MCMC sampling algorithms, are discussed. Finally, the choice of the parameters in SDIS is addressed and the coefficient of variation (CV) of SDIS estimator is analyzed. In Section 4, several benchmarks are used to assess the performance of our method, and some conclusions are drawn in Section 5.

## 2. Background

To solve the reliability analysis problem in Eq. (1), it is common to perform an isoprobabilistic transformation $\boldsymbol{U} = \boldsymbol{T}(\boldsymbol{X})$ to transform the random variables $\boldsymbol{X}$ to an equivalent space, the $\boldsymbol{U}$-space, where $\boldsymbol{U}$ is a vector of independent standard normal random variables [33]. As a result, the probability of failure can be expressed as

$$P_f = \int_{G(\boldsymbol{u})\leq 0} \varphi_n(\boldsymbol{u})\, d\boldsymbol{u} = \int_{\mathbb{R}^n} I(G(\boldsymbol{u}) \leq 0)\varphi_n(\boldsymbol{u})\, d\boldsymbol{u}, \tag{2}$$

where $\varphi_n(\boldsymbol{u})$ represents the $n$-variate standard normal PDF and $G(\boldsymbol{u}) = g\left(\boldsymbol{T}^{-1}(\boldsymbol{u})\right)$ denotes the transformed LSF in $\boldsymbol{U}$-space.

The integral in Eq. (2) can be estimated by standard MCS, through generating $N$ independent samples $\{\boldsymbol{u}_j\}_{j=1}^{N}$ following $\varphi_n(\boldsymbol{u})$, and taking the sample mean of the indicator function, i.e.,

$$P_f \approx \hat{P}_f = \frac{1}{N}\sum_{j=1}^{N} I\big(G(\boldsymbol{u}_j) \le 0\big). \quad (3)$$

The MCS estimator is unbiased, and the corresponding CV is given by

$$\delta_{\hat{P}_f} = \sqrt{\frac{1-P_f}{NP_f}}. \quad (4)$$

The CV in Eq. (4) is generally used to measure the accuracy of $\hat{P}_f$. For problems with $P_f$ in the order of $10^{-k}$, MCS requires approximately $N = 10^{k+2}$ samples to achieve an accuracy of $\delta_{\hat{P}_f} = 0.1$.

IS aims at improving the computational efficiency of the MCS estimator through reformulating the integral in Eq. (2) as

$$P_f = \int_{\mathbb{R}^n} I(G(\boldsymbol{u}) \le 0)\frac{\varphi_n(\boldsymbol{u})}{h(\boldsymbol{u})}h(\boldsymbol{u})d\boldsymbol{u} = E_h[I(G(\boldsymbol{U}) \le 0)W(\boldsymbol{U})], \quad (5)$$

where $h(\boldsymbol{u})$ is the IS density function, $E_h[\cdot]$ is the expectation operator with respect to density $h(\boldsymbol{u})$ and $W(\boldsymbol{u}) = \varphi_n(\boldsymbol{u})/h(\boldsymbol{u})$ is known as the importance weight function. The IS estimator of $P_f$ is obtained by generating $N$ independent samples $\{\boldsymbol{u}_j\}_{j=1}^{N}$ from $h(\boldsymbol{u})$ and approximating the expectation of Eq. (5) by a sample mean

$$P_f \approx \hat{P}_f = \frac{1}{N}\sum_{j=1}^{N} I\big(G(\boldsymbol{u}_j) \le 0\big)W(\boldsymbol{u}_j). \quad (6)$$

IS can reduce the CV of $\hat{P}_f$ dramatically compared to crude MCS, provided that a proper IS density function $h(\boldsymbol{u})$ is used. However, choice of a suitable IS density function is a nontrivial task. Although the optimal IS density function in theory is given by $h_{\text{opt}}(\boldsymbol{u}) = I(G(\boldsymbol{u}) \le 0)\varphi_n(\boldsymbol{u})/P_f$, it cannot be used in practice since it requires knowledge of $P_f$ and the location of the failure domain. Therefore, several adaptive methods, such as SuS [15] and SIS [21], define a family of intermediate IS density functions to successively approximate $h_{opt}(\boldsymbol{u})$. These methods employ Markov chain Monte Carlo (MCMC) algorithms to generate samples approximately distributed according to $h_{opt}(\boldsymbol{u})$.

The probability of failure in Eq. (2) can be also estimated in polar coordinates with DS [30, 31], which takes advantage of the polar representation of the coordinates of the $\boldsymbol{U}$-space, namely,

$$\boldsymbol{U} = R\boldsymbol{A}, \quad (7)$$

where $R^2 = \|\boldsymbol{U}\|_2^2$ follows the $\chi^2$-distribution with $n$ degrees of freedom, $\boldsymbol{A} = \boldsymbol{U}/\|\boldsymbol{U}\|_2$ is independent of $R$ and follows the uniform distribution on the $n$-dimensional hypersphere $\mathbb{S}^{n-1}$. The sample pair $\{r, \boldsymbol{a}\}$ represents the polar coordinates (radius and direction) of a corresponding standard normal sample $\boldsymbol{u}$. In the polar coordinates, the probability of failure can be reformulated by DS as

$$
\begin{aligned}
P_f &= \int_{G(\boldsymbol{u})\le 0} \varphi_n(\boldsymbol{u})\, d\boldsymbol{u} \\
&= \int_{\mathbb{S}^{n-1}} \int_{G(r\boldsymbol{a})\le 0} f_{R|\boldsymbol{A}}(r)\, dr\, f_{\boldsymbol{A}}(\boldsymbol{a}) d\boldsymbol{a} \\
&= \int_{\mathbb{S}^{n-1}} \Pr(G(r\boldsymbol{a}) \le 0)\, f_{\boldsymbol{A}}(\boldsymbol{a}) d\boldsymbol{a} \\
&= \int_{\mathbb{S}^{n-1}} \left[1 - F_{\chi_n^2}\left(r^2(\boldsymbol{a})\right)\right] f_{\boldsymbol{A}}(\boldsymbol{a}) d\boldsymbol{a},
\end{aligned} \tag{8}
$$

where $f_{\boldsymbol{A}}(\boldsymbol{a})$ denotes the PDF of the directional random vector $\boldsymbol{A}$, $f_{R|\boldsymbol{A}}(r)$ is the conditional PDF of radius $R$ along the direction $\boldsymbol{A}$, $F_{\chi_n^2}(\cdot)$ represents the cumulative density function (CDF) of the $\chi^2$-distribution with $n$ degrees of freedom, and $r(\boldsymbol{a})$ is the root of $G(r\boldsymbol{a}) = 0$ on direction $\boldsymbol{a}$.

To estimate the integral in Eq. (8), the first step of DS consists in generating directional vectors $\{\boldsymbol{a}_j, j = 1, \dots, N\}$ uniformly on the unit hyper-sphere $\mathbb{S}^{n-1}$. For each direction $\boldsymbol{a}_j$, the root $r(\boldsymbol{a}_j)$ of the limit state $G(r\boldsymbol{a}) = 0$ is found by application of a line-search method. The DS estimate of the failure probability is then given by

$$
P_f \approx \hat{P}_f = \frac{1}{N} \sum_{j=1}^{N} \left[1 - F_{\chi_n^2}\left(r^2(\boldsymbol{a}_j)\right)\right]. \tag{9}
$$

The DS estimator in Eq. (9) is generally inefficient for high-dimensional problems since for large $n$ the number of directions needed to sufficiently explore the unit hypersphere becomes significant and most of these direction point to the non-important regions [34]. This problem can be alleviated by the DIS method, which reformulates Eq. (8) as

$$
P_f = \int_{\mathbb{S}^{n-1}} \left[1 - F_{\chi_n^2}\left(r^2(\boldsymbol{a})\right)\right] \frac{f_{\boldsymbol{A}}(\boldsymbol{a})}{h(\boldsymbol{a})} h(\boldsymbol{a}) d\boldsymbol{a} = E_h\left(\left[1 - F_{\chi_n^2}\left(r^2(\boldsymbol{A})\right)\right] W(\boldsymbol{A})\right), \tag{10}
$$

where $h(\boldsymbol{a})$ is the DIS density function, and $W(\boldsymbol{a}) = f_{\boldsymbol{A}}(\boldsymbol{a})/h(\boldsymbol{a})$ represents the importance weight function. Through generating $N$ independent samples $\{\boldsymbol{a}_j, j = 1, \dots, N\}$ from $h(\boldsymbol{a})$, DIS approximates $P_f$ as

$$
P_f \approx \hat{P}_f = \frac{1}{N} \sum_{j=1}^{N} \left[1 - F_{\chi_n^2}\left(r^2(\boldsymbol{a}_j)\right)\right] W(\boldsymbol{a}_j). \tag{11}
$$

The theoretically optimal DIS density function $h_{\text{opt}}(\boldsymbol{a})$ is given by

$$
h_{\text{opt}}(\boldsymbol{a}) = \frac{\left[1 - F_{\chi_n^2}\left(r^2(\boldsymbol{a})\right)\right] f_{\boldsymbol{A}}(\boldsymbol{a})}{P_f}. \tag{12}
$$

The optimal DIS function results in a zero-variance DIS estimator. However, similar with IS, $h_{\text{opt}}(\boldsymbol{a})$ cannot be implemented directly due to lack of knowledge about $P_f$ , and it is a challenging task to select a proper $h(\boldsymbol{a})$ for application in DIS. Some suboptimal choices of $h(\boldsymbol{a})$ are discussed in [31, 35, 36].

## 3. Methodology

In this section, we introduce the proposed SDIS method for reliability analysis. SDIS defines a sequence of reliability problems, such that the first problem can be solved with crude MCS using a small number of samples and the last problem corresponds to the sought reliability problem. Each intermediate problems is solved using DIS with samples drawn from the optimal DIS density of the previous problem.

We first define the family of auxiliary failure probabilities that define the sequence of reliability problems, by magnifying the input variability. This is followed by a detailed description of the rationale of SDIS. Thereafter, a resampling scheme and two MCMC algorithms are presented for sampling from the optimal DIS densities of the intermediate problems in SDIS. Finally, the choice of the parameters in SDIS as well as the CV of the SDIS estimator are discussed.

### *3.1 Auxiliary failure probabilities*

In problems with a small failure probability $P_f$, the failure domain is generally far from the high probability mass region of $\varphi_n(\boldsymbol{u})$. As a result, crude MCS requires lots of samples (on average $P_f^{-1}$) to obtain a single failure sample and a multitude thereof to obtain an accurate probability estimate. To improve the sampling efficiency, one can amplify the standard deviation of all the input variables [37-39], resulting in the following auxiliary failure probability

$$P_\sigma = \int_{\mathbb{R}^n} I(G(\boldsymbol{u}) \leq 0) f_\sigma(\boldsymbol{u}) d\boldsymbol{u}. \tag{13}$$

where $f_\sigma(\boldsymbol{u}) = \prod_{l=1}^{n} f_\sigma(u_l)$, and

$$f_\sigma(u) = \frac{1}{\sqrt{2\pi}\sigma} \exp\left(-\frac{u^2}{2\sigma^2}\right), \tag{14}$$

in which $\sigma(\sigma > 1)$ represents the standard deviation (magnification factor) of every random variable $U_l (l = 1, \dots, n)$ . Compared to the original PDF $\varphi_n(\boldsymbol{u})$ , more probability mass of the inflated PDF $f_\sigma(\boldsymbol{u})$ resides in the failure domains, and thus one can generate failure samples efficiently by sampling from $f_\sigma(\boldsymbol{u})$ .

Alternatively, $P_\sigma$ can be equivalently expressed as the following integral

$$P_\sigma = \int_{\mathbb{R}^n} I(G(\sigma\boldsymbol{u}) \leq 0) \varphi_n(\boldsymbol{u}) d\boldsymbol{u}, \tag{15}$$

where $G(\sigma\boldsymbol{u})$ is an auxiliary LSF [37]. Compared to $G(\boldsymbol{u})$, the failure domain of the auxiliary LSF $G(\sigma\boldsymbol{u})$ is closer to the high probability mass of $\varphi_n(\boldsymbol{u})$, and thus failure samples, satisfying $G(\sigma\boldsymbol{u}) \leq 0$, can be efficiently drawn from $\varphi_n(\boldsymbol{u})$[37, 40].

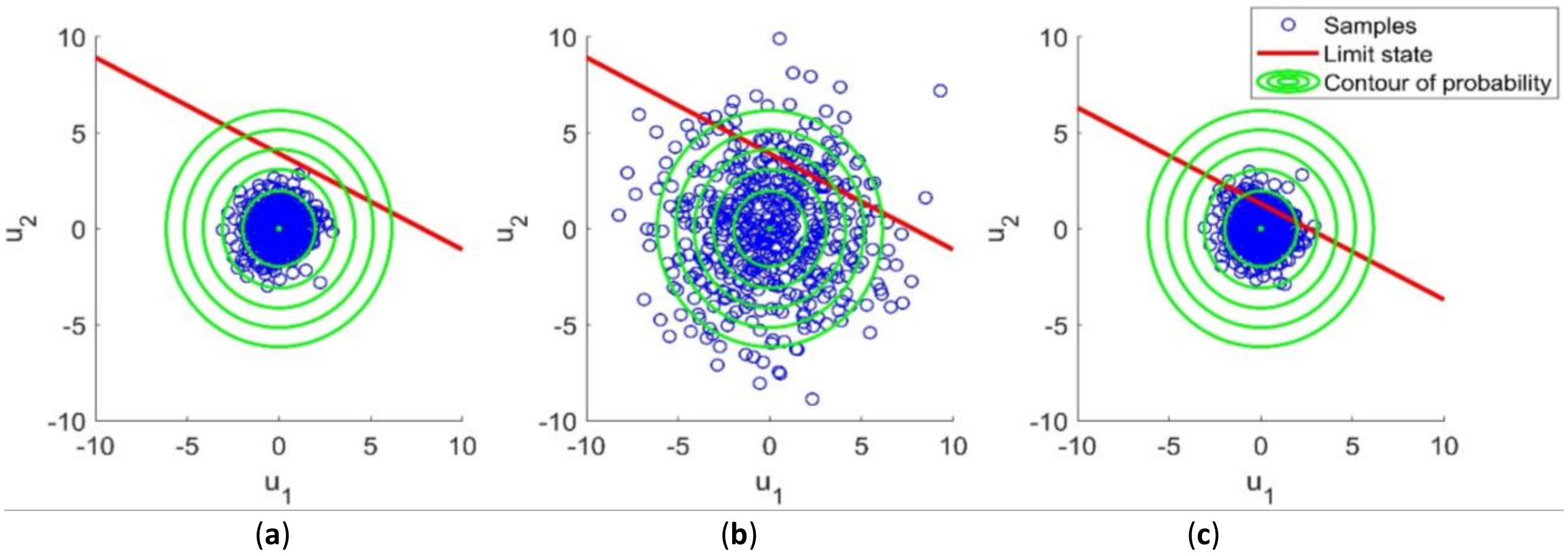

**Fig. 1**. Comparisons of sampling efficiency for exploring the failure domain of a linear LSF with 500 random samples. **(a)** Original reliability problem in standard normal space. (**b**) Modified reliability problem with samples drawn from inflated PDF $f_3(\boldsymbol{u})$. (**c**) Modified reliability problem with auxiliary LSF $G(3\boldsymbol{u})$.

In Fig. 1, a two-dimensional example is used to illustrate the basic principle of the modified reliability problem of Eq. (11) and the equivalent problem of Eq. (13). It is observed from Fig. 1(a) that all the 500 random samples reside in the safety domain in the standard normal space. On the contrary, by tripling the standard deviation of the two inputs, a large group of failure samples is obtained, as depicted in Fig. 2(b). In Fig. 1(c), we draw samples from standard normal PDF $\varphi_n(\boldsymbol{u})$, and evaluate the auxiliary LSF $G(3\boldsymbol{u})$. One can see that a large number of failure samples is also obtained in this case.

With $\sigma$ approaching 1, the inflated PDF $f_\sigma(\boldsymbol{u})$ and the auxiliary LSF $G(\sigma\boldsymbol{u})$ approach the original standard normal PDF $\varphi_n(\boldsymbol{u})$ and the true LSF $G(\boldsymbol{u})$ respectively, and the auxiliary failure probability $P_\sigma$ approaches the sought probability $P_f$.

*3.2 Sequential directional importance sampling*

In this subsection, we describe the basic principle of the SDIS algorithm for reliability analysis. As mentioned previously, SDIS is a variant of SIS. In SIS [21], a series of intermediate IS density functions is introduced by replacing the indicator function in the optimal IS density with a sequence of smooth approximation thereof. The normalizing constant of each intermediate density is estimated with IS using samples from the previous density. Consequently, SIS converts a rare event estimation problem into a sequence of estimation problems, whereby each problem can be efficiently estimated with a small number of samples. SDIS follows the same principle, i.e., it introduces a series of intermediate IS density functions, but constructs the sequence by replacing the original LSF in the optimal IS density with the family of auxiliary LSFs introduced in Section 3.1. The normalizing constant of each intermediate density or, equivalently, the failure probability of the reliability problem with corresponding auxiliary LSF, is estimated using samples from the previous IS density. Thereby, estimation of the first probability is performed in cartesian coordinates, and all the subsequent ones are computed in polar coordinates.

In the first step of SDIS, we magnify the standard deviations of all inputs by $\sigma_1$ times, and thus the initial auxiliary failure probability $P_{\sigma_1}$ can be expressed as

$$P_{\sigma_1} = \int_{\mathbb{R}^n} I(G(\sigma_1 \boldsymbol{u}) \leq 0)\ \varphi_n(\boldsymbol{u}) d\boldsymbol{u} = \int_{\mathbb{S}^{n-1}} \left[1 - F_{\chi_n^2}\left(r_1^2(\boldsymbol{a})\right)\right] f_{\boldsymbol{A}}(\boldsymbol{a}) d\boldsymbol{a}, \tag{16}$$

where $r_1(\boldsymbol{a})$ is the root of $G(\sigma_1 r \boldsymbol{a}) = 0$ on direction $\boldsymbol{a}$. Then, $P_{\sigma_1}$ is estimated with MCS in cartesian coordinates, namely

$$P_{\sigma_1} \approx \hat{P}_{\sigma_1} = \frac{1}{N} \sum_{j=1}^{N} I\left(G\left(\sigma_1 \boldsymbol{u}_j\right) \leq 0\right) = \frac{N_0}{N} \tag{17}$$

where $\boldsymbol{u}_j (j = 1, \ldots, N)$ are random samples independently drawn from $\varphi_n(\boldsymbol{u})$ and $N_0$ is the number of failure samples.

To guarantee the accuracy of $P_{\sigma_1}$, we increase the MCS population sequentially until $N_0 (N_0 = 100,200, \ldots)$ failure samples are obtained, which will guarantee that the CV of $P_{\sigma_1}$ is $\lesssim 1/\sqrt{N_0}$. Since the standard deviation of every input is magnified, these $N_0$ failure samples can be generated efficiently. Every failure sample drawn in this step defines an important direction $\boldsymbol{a}_j = \boldsymbol{u}_j / \left\|\boldsymbol{u}_j\right\|_2 (j = 1, \ldots, N_0)$, which points towards the failure domain.

Then, the second auxiliary failure probability $P_{\sigma_2}$ can be estimated with DIS as

$$\begin{aligned} P_{\sigma_2} &= \int_{\mathbb{R}^n} I(G(\sigma_2 \boldsymbol{u}) \leq 0) \varphi_n(\boldsymbol{u}) d\boldsymbol{u} \\ &= \int_{\mathbb{S}^{n-1}} \left[1 - F_{\chi_n^2}\left(r_2^2(\boldsymbol{a})\right)\right] f_{\boldsymbol{A}}(\boldsymbol{a}) d\boldsymbol{a} \\ &= P_{\sigma_1} \int_{\mathbb{S}^{n-1}} \left[\frac{1 - F_{\chi_n^2}\left(r_2^2(\boldsymbol{a})\right)}{1 - F_{\chi_n^2}\left(r_1^2(\boldsymbol{a})\right)}\right] \frac{\left[1 - F_{\chi_n^2}\left(r_1^2(\boldsymbol{a})\right)\right] f_{\boldsymbol{A}}(\boldsymbol{a})}{P_{\sigma_1}} d\boldsymbol{a} \\ &= P_{\sigma_1} E_{h_1}[W_1(\boldsymbol{A})] \end{aligned} \tag{18}$$

where $\sigma_2 (\sigma_2 \leq \sigma_1)$ is the magnification factor of input standard deviation in the second step, $r_2(\boldsymbol{a})$ is the root of $G(\sigma_2 r \boldsymbol{a}) = 0$ on direction $\boldsymbol{a}$, $h_1(\boldsymbol{a}) = \left[1 - F_{\chi_n^2}\left(r_1^2(\boldsymbol{a})\right)\right] f_{\boldsymbol{A}}(\boldsymbol{a}) / P_{\sigma_1}$ is the IS density function of $\boldsymbol{a}$, and $W_1(\boldsymbol{a}) = \left(1 - F_{\chi_n^2}\left(r_2^2(\boldsymbol{a})\right)\right) \Big/ \left(1 - F_{\chi_n^2}\left(r_1^2(\boldsymbol{a})\right)\right)$ is the importance weight.

This procedure is repeated for every subsequent step, and $P_f$ can be finally reformulated as

$$
\begin{aligned}
P_f &= \int_{\mathbb{R}^n} I(G(\sigma_k \boldsymbol{u}) \le 0)\varphi_n(\boldsymbol{u})d\boldsymbol{u} \\
&= \int_{\mathbb{S}^{n-1}} \left[1 - F_{\chi_n^2}\left(r_k^2(\boldsymbol{a})\right)\right] f_{\boldsymbol{A}}(\boldsymbol{a})d\boldsymbol{a} \\
&= P_{\sigma_{k-1}} \int_{\mathbb{S}^{n-1}} \left[\frac{1 - F_{\chi_n^2}\left(r_k^2(\boldsymbol{a})\right)}{1 - F_{\chi_n^2}\left(r_{k-1}^2(\boldsymbol{a})\right)}\right] \frac{\left[1 - F_{\chi_n^2}\left(r_{k-1}^2(\boldsymbol{a})\right)\right] f_{\boldsymbol{A}}(\boldsymbol{a})}{P_{\sigma_{k-1}}} d\boldsymbol{a} \\
&\vdots \\
&= P_{\sigma_1} \prod_{i=1}^{k-1} S_i
\end{aligned} \tag{19}
$$

where $S_i = E_{h_i}[W_i(\boldsymbol{A})]$, $1 = \sigma_k \le \sigma_{k-1} \le \cdots \le \sigma_1$, $r_i(\boldsymbol{a})$ is the root of $G(\sigma_i r\boldsymbol{a}) = 0$ on direction $\boldsymbol{a}$,

$$h_i(\boldsymbol{a}) = \frac{\left[1 - F_{\chi_n^2}\left(r_i^2(\boldsymbol{a})\right)\right] f_{\boldsymbol{A}}(\boldsymbol{a})}{P_{\sigma_i}} \tag{20}$$

is the IS density function of $\boldsymbol{a}$, and

$$W_i(\boldsymbol{a}) = \frac{1 - F_{\chi_n^2}\left(r_{i+1}^2(\boldsymbol{a})\right)}{1 - F_{\chi_n^2}\left(r_i^2(\boldsymbol{a})\right)} \tag{21}$$

is the importance weight in each step.

The roots $r_{i+1}(\boldsymbol{a})$ and $r_i(\boldsymbol{a})$ of $G(\sigma_{i+1} r\boldsymbol{a}) = 0$ and $G(\sigma_i r\boldsymbol{a}) = 0$ respectively satisfy the following relationship

$$r_{i+1}(\boldsymbol{a}) = r_i(\boldsymbol{a})\sigma_i/\sigma_{i+1}. \tag{22}$$

With Eq. (19), we decompose the original failure probability in Eq. (8) into a series of integrals, in which $P_{\sigma_1}$ is estimated with MCS following Eq. (15), and the subsequent integrals $S_i = E_{h_i}[W_i(\boldsymbol{A})]$ can be estimated with DIS efficiently using samples from $h_i(\boldsymbol{a})$, resulting in

$$S_i \approx \hat{S}_i = \frac{1}{N_0}\sum_{j=1}^{N_0} W_i(\boldsymbol{a}_j) = \frac{1}{N_0}\sum_{j=1}^{N_0} \frac{1 - F_{\chi_n^2}\left(r_{i+1}^2(\boldsymbol{a}_j)\right)}{1 - F_{\chi_n^2}\left(r_i^2(\boldsymbol{a}_j)\right)}, \tag{23}$$

where $\boldsymbol{a}_j (j = 1, \ldots, N_0)$ are samples drawn from $h_i(\boldsymbol{a})$. We note that evaluation of the importance weight in Eq. (23) for each sample $\boldsymbol{a}_k$ requires solving a single line search problem, e.g., for $r_i(\boldsymbol{a}_j)$, due to Eq. (22). Samples from $h_i(\boldsymbol{a})$ can be obtained starting from samples following $h_{i-1}(\boldsymbol{a})$ through a resample-move scheme. First, samples $\{\boldsymbol{a}_j, j = 1, \ldots, N_0\}$ from $h_{i-1}(\boldsymbol{a})$ are resampled according to the weights $W_{i-1}(\boldsymbol{a}_j)$, to obtain samples that are asymptotically distributed according to $h_i(\boldsymbol{a})$. Thereafter, the samples are moved through an MCMC step with invariant distribution $h_i(\boldsymbol{a})$. The resampling and MCMC steps are discussed in the following sections.

3.3 Resampling scheme

In the SDIS procedure, one needs to generate samples of $\boldsymbol{a}$ following $h_i(\boldsymbol{a})$ . To this end, we first generate samples $\boldsymbol{u}$ in Cartesian coordinates following

$$\tilde{h}_i(\boldsymbol{u}) = \frac{I(G(\sigma_i \boldsymbol{u}) \leq 0)\varphi_n(\boldsymbol{u})}{P_{\sigma_i}}. \tag{24}$$

Then, we transform the samples $\boldsymbol{u}$ drawn from $\tilde{h}_i(\boldsymbol{u})$ in Cartesian coordinates to directional samples $\boldsymbol{a}$ from $h_i(\boldsymbol{a}) = \left[1 - F_{\chi_n^2}\left(r_i^2(\boldsymbol{a})\right)\right] f_{\boldsymbol{A}}(\boldsymbol{a})/P_{\sigma_i}$ in polar coordinates by setting $\boldsymbol{a} = \boldsymbol{u}/\|\boldsymbol{u}\|_2$.

To obtain samples from the density of Eq. (24), we resample from existing samples available from the previous sampling step. Specifically, to estimate $P_{\sigma_2}$ , samples following $\tilde{h}_1(\boldsymbol{u}) = I(G(\sigma_i \boldsymbol{u}) \leq 0)\varphi_n(\boldsymbol{u})/P_{\sigma_1}$ can be obtained directly by selecting the $N_0$ failure samples $\boldsymbol{u}_j (j = 1, \ldots, N_0$ ) from the MCS population in the first step. By setting $\boldsymbol{a}_j = \boldsymbol{u}_j/\|\boldsymbol{u}_j\|_2$, we obtain $N_0$ samples of $\boldsymbol{a}$ following $h_1(\boldsymbol{a}) = \left[1 - F_{\chi_n^2}\left(r_1^2(\boldsymbol{a})\right)\right] f_{\boldsymbol{A}}(\boldsymbol{a})/ P_{\sigma_1}$ perfectly. After finding the root $r_1(\boldsymbol{a}_j)(j = 1, \ldots, N_0$ ) of $G(\sigma_1 r\boldsymbol{a}) = 0$ along each direction, $P_{\sigma_2}$ can be estimated as

$$P_{\sigma_2} \approx \hat{P}_{\sigma_2} = \hat{P}_{\sigma_1} \frac{1}{N_0} \sum_{j=1}^{N_0} \frac{\left(1 - F_{\chi_n^2}\left(r_2^2(\boldsymbol{a}_j)\right)\right)}{\left(1 - F_{\chi_n^2}\left(r_1^2(\boldsymbol{a}_j)\right)\right)}, \tag{25}$$

where $r_2(\boldsymbol{a}_j)$ is the root of $G(\sigma_2 r\boldsymbol{a}_j) = 0$ on direction $\boldsymbol{a}_j$, and it can be computed as $r_2(\boldsymbol{a}_j) = \sigma_1 r_1(\boldsymbol{a}_j)/\sigma_2$ without additional computational cost, as depicted in Fig. 2(a).

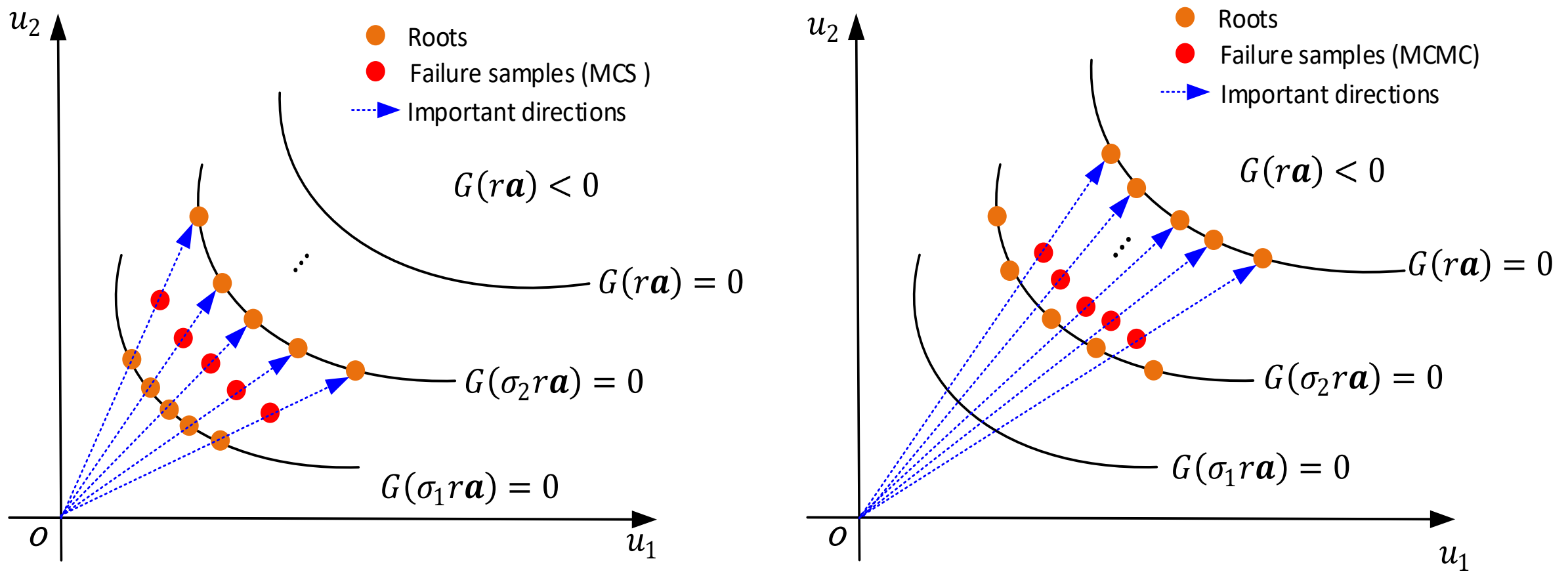

(a) (b)

**Fig. 2**. Basic principle of SDIS; (a) Initial step of SDIS for exploiting failure domains globally with MCS; (b) Subsequent steps of SDIS for exploring failure domains sequentially with MCMC.

In the subsequent steps, we select the initial samples following $\tilde{h}_i(\boldsymbol{u})$ $(i > 1)$ asymptotically by applying the following scheme.

1. Resampling of direction: Select $N_0$ samples $\boldsymbol{a}_j (j = 1, \dots, N_0)$ of direction (with replacement) following $h_{i-1}(\boldsymbol{a})$ randomly with probability assigned to each sample proportional to $W_{i-1}(\boldsymbol{a})$.

2. Resampling of radius: For each direction $\boldsymbol{a}_j (j = 1, \dots, N_0)$ obtained in step 1, generate a radius sample $r(\boldsymbol{a}_j)$ following the truncated $\chi$-distribution with $n$ degrees of freedom truncated given $r(\boldsymbol{a}_j) > r_i(\boldsymbol{a}_j)$.

By transforming the obtained samples pair $\{r(\boldsymbol{a}_j), \boldsymbol{a}_j\} (j = 1, \dots, N_0)$ into Cartesian coordinates through $\boldsymbol{u}_j = r(\boldsymbol{a}_j)\boldsymbol{a}_j$, we can obtain $N_0$ samples from $\tilde{h}_i(\boldsymbol{u})$. To further obtain samples in the high probability density region of $\tilde{h}_i(\boldsymbol{u})$, we apply MCMC sampling algorithms, which are described in the following subsection.

*3.4 MCMC sampling*

For each sample $\boldsymbol{u}_j (j = 1, \dots, N_0)$ obtained from the resampling scheme introduced in subsection 3.3, MCMC sampling is used to simulate states of a Markov chain of length $l + 1$ with stationary distribution equal to the target distribution $\tilde{h}_i(\boldsymbol{u})$. Specifically, we adopt the popular Metropolis-Hastings (M-H) algorithm, which is performs the following steps:

1. Select a seed $\boldsymbol{u}_t$, and set $t = 0$.
2. Generate a random candidate state $\boldsymbol{u}'$ from a proposal $q(\boldsymbol{u}'|\boldsymbol{u}_t)$, and compute the corresponding response $y = G(\sigma_i \boldsymbol{u}')$.
3. Calculate the acceptance probability

$$\alpha = I(G(\sigma_i \boldsymbol{u}') \leq 0) \min\left(1, \frac{\varphi_n(\boldsymbol{u}')q(\boldsymbol{u}_t|\boldsymbol{u}')}{\varphi_n(\boldsymbol{u}_t)q(\boldsymbol{u}'|\boldsymbol{u}_t)}\right). \tag{26}$$

4. Generate a uniform random number $w \in [0,1]$, and determine the next state as

$$\boldsymbol{u}_{t+1} = \begin{cases} \boldsymbol{u}, & \alpha \geq w, \\ \boldsymbol{u}_t, & \alpha < w. \end{cases} \tag{27}$$

5. Set $t = t + 1$, and go back to step 2 until the target length $l$ is reached.

Using above M-H algorithm, we generate $N_0$ Markov chains of length $l + 1$, thus we have $(l + 1)N_0$ failure samples in total (including the seeds), as shown in Fig. 2(b). To reduce the effect of the transient period of the simulated Markov chain, a burn-in period is applied through discarding the first $l$ samples from each Markov chain. Hence, the last state of each Markov chain is kept, which is expected to follow $\tilde{h}_i(\boldsymbol{u})$ perfectly. In this work, we suggest setting $l = 5$ to ensure the accuracy of the SDIS estimator (see the discussion in Section 4). Finally, by setting $\boldsymbol{a} = \boldsymbol{u}/\|\boldsymbol{u}\|_2$, we obtain $N_0$ samples following $h_i(\boldsymbol{a})$.

The performance of the above M-H algorithm depends on the choice of proposal distribution $q(\boldsymbol{u}'|\boldsymbol{u}_t)$. In the current paper, two proposal distributions are considered. The first one leads to an independent sampler that is very efficient for low to moderate dimensional problems, and the second one is based on sampling from a conditional normal distribution in standard normal space, which is efficient for both low and high-dimensional problems.

*3.4.1 Independent M-H (IM-H) algorithm*

The first proposal distribution we considered is a Gaussian mixture model which is independent of the current state $\boldsymbol{u}_t$, and is defined as [21]

$$\pi(\boldsymbol{u}') = \sum_{j=1}^{K} p_j \varphi\left(\boldsymbol{u}' \middle| \boldsymbol{\mu}_j, \boldsymbol{\Sigma}_j\right), \tag{28}$$

where $p_j$ is the weight, $\boldsymbol{\mu}_j$ and $\boldsymbol{\Sigma}_j$ are the mean vector and covariance matrix of the $j$-th Gaussian distribution $\varphi(\cdot)$. The parameters of the Gaussian mixture model can be estimated with the so-called expectation-maximization (EM) algorithm [41] using the samples obtained after the resampling step. We note that to obtain a reliable fit of the Gaussian mixture model, a larger number of samples can be generated conditional on each direction through generating multiple radii from the truncated $\chi$-distribution (step 2 in the resampling algorithm of Section 3.4). With this proposal, the acceptance probability becomes

$$\alpha = I(G(\sigma_i \boldsymbol{u}') \leq 0) \min\left(1, \frac{\varphi_n(\boldsymbol{u}')\pi(\boldsymbol{u}_t)}{\varphi_n(\boldsymbol{u}_t)\pi(\boldsymbol{u}')}\right). \tag{29}$$

The independent M–H algorithm presented in this subsection is effective for problems with low and medium dimension. For high-dimensional problems, the ratio $\varphi_n(\boldsymbol{u}')/\varphi_n(\boldsymbol{u}_t)$ appearing in the acceptance probability $\alpha$ tends to be extremely small, which undermines the performance of this method [21]. To avoid this issue, one can use the conditional sampling M-H algorithm presented in the subsequent subsection.

*3.4.2 Conditional sampling M-H (CSM-H) algorithm*

The second proposal distribution we used in this paper is the normal distribution conditioned on the current state (conditional M-H algorithm) [42, 43], i.e., $\boldsymbol{u}' = \sqrt{1-\beta^2}\boldsymbol{u}_t + \beta\xi_t$ with $\xi_t \sim N(0,1)$ and $\beta \in (0,1)$. Based on this proposal, the acceptance probability can be rewritten as

$$\alpha = I(G(\sigma_i \boldsymbol{u}') \leq 0). \tag{30}$$

The acceptance probability of Eq. (30) only depends on the indicator function of the candidate state and is thus suitable for high dimensional problems. The parameter $\beta$ in the proposal can be chosen freely. A small $\beta$ implies that the candidate state $\boldsymbol{u}'$ is close to current state $\boldsymbol{u_t}$, and the correlation between states will be strong. On the contrary, a large $\beta$ implies that the candidate state $\boldsymbol{u}'$ is far away from the current state, and this will lead to many rejected candidates. In the current paper, we use the method proposed in [43] to adjust $\beta$ adaptively so as to guarantee that the acceptance probability $\alpha$ remains close to the optimal value 44%. Note that the performance of this algorithm is independent of the input dimensionality.

*3.5 Choice of* $\sigma$

In SDIS, one needs to choose an initial value of the magnification factor $\sigma_1$ to start the simulation procedure. The initial value of $\sigma_1$ should be large enough such that the first step of SDIS is able to sufficiently explore the parameter space and capture all important regions of the failure domains rapidly. However, for very large $\sigma_1$, many intermediate steps are required to obtain a convergent estimation of failure probability, which will in turn undermine the performance of SDIS. Based on our experience and the literature [37], we suggest setting $\sigma_1$ =3 (see the discussion in Section 4). To ensure the accuracy of the integral $S_i$ in each step, one can select the subsequent magnification factors $\sigma_i$ adaptively such that the sample CV $\hat{\delta}_{W_i}$ of $\{W_i(\boldsymbol{a}_j), j = 1, \ldots, N_0\}$ takes a target value. To this end, the magnification factors can be determined by the following optimization problem [21]

$$\sigma_i = \underset{\sigma_i \in [1, \sigma_{i-1}]}{\operatorname{argmin}} \left| \hat{\delta}_{W_i} - \delta_{\text{target}} \right|, (i \geq 2), \tag{31}$$

where $\delta_{\text{target}}$ is the target value of the CV. Note that above optimization problem does not require additional model evaluations.

We now discuss the effectiveness of this adaptive approach. To this end, the CV of the SDIS estimator is analyzed. We assume that the samples following $h_i(\boldsymbol{a})$ are independent from samples from other sampling levels $h_k(\boldsymbol{a})(k \neq i)$, and the samples at each level drawn from $h_i(\boldsymbol{a})$ are independent from each other. These assumptions are not very restrictive as their satisfaction can be guaranteed through applying a sufficient burn-in period in the MCMC step. Then, the CV of the SDIS estimator can be expressed as [44]

$$\delta_{\hat{P}_f} = \sqrt{\delta^2_{\hat{P}_{\sigma_1}} + \sum_{i=1}^{k-1} \delta^2_{\hat{S}_i}}, \tag{32}$$

where $\delta_{\hat{P}_1}$ and $\delta_{\hat{S}_i}$ are the CV of $\hat{P}_{\sigma_1}$ and $\hat{S}_i$ respectively, and

$$\delta_{\hat{S}_i}^2 = \frac{1}{N_0}\delta_{W_i}^2, \tag{33}$$

where $\delta_{W_i}$ is the CV of $W_i(\boldsymbol{a})$. Since $\hat{P}_{\sigma_1}$ is estimated with MCS by generating $N_0$ failure samples, its CV is $\delta_{\hat{P}_1} \lesssim 1/\sqrt{N_0}$ according to Eq. (4). Therefore, bounding $\delta_{\hat{P}_f}$ with threshold $\varepsilon$ is equivalent to bounding $\delta_{w_i}$ in each intermediate step, namely,

$$\delta_{\hat{P}_f} \leq \varepsilon \leftrightarrow \delta_{W_i} \leq \sqrt{\frac{\left(\varepsilon^2 - \delta_{\hat{P}_{\sigma_1}}^2\right)N_0}{k-1}} \leq \sqrt{\frac{N_0\varepsilon^2 - 1}{k-1}}. \tag{34}$$

Through solving the optimization problem in Eq. (29), $\hat{\delta}_{W_i}$ will adhere to $\delta_{\text{target}}$, and this allow us to control the CV of the SDIS estimator. In this paper, we set $\delta_{\text{target}} = 1.5$ as proposed in [44].

*3.6 Summary of SDIS algorithm*

The SDIS algorithm for rare event estimation is summarized as follows:

1. Initialize: Set the initial magnification factor $\sigma_1 = 3$, the length of each Markov chain $l = 5$ and the number of important directions per level $N_0$ (here we suggest setting $N_0 = 100$). Set $k = 1$.

2. Compute $\hat{P}_{\sigma_1}$ with Eq. (17) using MCS by generating $N_0$ failure samples.

3. Find the root of $G(\sigma_k\boldsymbol{u}) = 0$ along the $N_0$ important directions, and evaluate the weights:

$$W_k(\boldsymbol{a}_j) = \frac{1 - F_{\chi_n^2}\left(r_{k+1}^2(\boldsymbol{a}_j)\right)}{1 - F_{\chi_n^2}\left(r_k^2(\boldsymbol{a}_j)\right)}, j = 1, \dots, N_0. \tag{35}$$

4. Determine parameter $\sigma_{k+1}$ by solving Eq. (31).

5. Compute the integral

$$\hat{S}_k = \frac{1}{N_0}\sum_{j=1}^{N_0} W_k(\boldsymbol{a}_j). \tag{36}$$

6. Resample: Select $N_0$ seeds $\boldsymbol{u}_j (j = 1, \dots, N_0)$ with the resampling method presented in Section 3.3.

7. Move: Run $N_0$ Markov chains independently of length $l + 1$ for each seed $\boldsymbol{u}_j$ obtained in step 6 with target distribution $\tilde{h}_i(\boldsymbol{u})$ by using one of the two MCMC algorithms presented in Section 3.4. Select the last state of each chain, and transform these $N_0$ samples to polar coordinate by $\boldsymbol{a} = \boldsymbol{u}/\|\boldsymbol{u}\|_2$.

8. Set $k = k + 1$, and go back to step 3 until $\sigma_k = 1$.

9. Estimate the failure probability as：

$$\hat{P}_f = \hat{P}_{\sigma_1} \prod_{i=1}^{k-1} \hat{S}_i. \tag{37}$$

## 4. Numerical examples

In this section, we assess the performance of the SDIS method with several benchmark examples, and the results are compared with the popular SuS method [15, 45] and with SIS. In SuS and SIS, the sample number per level is set as 1000. The intermediate failure probability of SuS is set to 0.1, and the target CV for selection of the intermediate distributions in SIS is set to $\delta_{\text{target}} = 1.5$. In SDIS, the "fsolve" function (using trust-region dogleg algorithm) in Matlab platform is used to find the root on each important direction. To evaluate the performance of each method comprehensively, the statistics of the estimates obtained by the different methods are computed with 300 independent simulation runs.

### 4.1. Four branch series system reliability problem

The first example is a series system [21, 38] defined by the following LSF

$$g(\boldsymbol{u}) = \min \left\{ \begin{matrix} 3 + 0.1(u_1 - u_2)^2 - (u_1 + u_2)/\sqrt{2} \\ 3 + 0.1(u_1 - u_2)^2 + (u_1 + u_2)/\sqrt{2} \\ (u_1 - u_2) + 6/\sqrt{2} \\ (u_2 - u_1) + 6/\sqrt{2} \end{matrix} \right\} + 2 \tag{38}$$

where $u_i (i = 1,2)$ are outcomes of standard normal distributed random variables.

The reference value of failure probability obtained by MCS with $10^9$ samples is $1.058 \times 10^{-5}$. In Fig. 3, a specific example run of the SDIS method is used to illustrate its performance. In this example run, by setting $\sigma_1 = 3$, the first auxiliary limit state surface $G(\sigma_1 \boldsymbol{u}) = 0$ is much closer to the high probability mass of $\varphi_2(\boldsymbol{u})$ compared to the original limit state surface, as shown in Fig. 3(a). We generate 517 samples following $\varphi_2(\boldsymbol{u})$ sequentially until 100 failure samples satisfying $G(\sigma_1 \boldsymbol{u}) \leq 0$ are obtained, and thus we have $P_{\sigma_1} \approx 0.1934$. As depicted in Fig. 3(a), these failure samples $\boldsymbol{u}_j$ define 100 important directions $\boldsymbol{a}_j = \boldsymbol{u}_j / \left\| \boldsymbol{u}_j \right\|_2$ $(k = 1, \ldots, 100)$ following $h_1(\boldsymbol{a}) = \left[1 - F_{\chi_n^2}\left(r_1^2(\boldsymbol{a})\right)\right] f_{\boldsymbol{A}}(\boldsymbol{a}) / P_{\sigma_1}$.

Using these 100 directional samples $\boldsymbol{a}_k$, we turn to compute the root $r_1(\boldsymbol{a})$ of $G(\sigma_1 r \boldsymbol{a}) = 0$ along these directions. Then, we determine $\sigma_2$ with Eq. (29), and its optimal value is 1 in this specific example. Consequently, the root $r_2(\boldsymbol{a})$ of $G(\sigma_2 r \boldsymbol{a}) = 0$ along these 100 important directions can be obtained by $r_2(\boldsymbol{a}_j) = 3 r_1(\boldsymbol{a}_j)$, as represented in Fig. 3(b). Note that true limit state surface coincides with the second auxiliary one in this example, and thus we obtain the estimation of $\hat{P}_f = \hat{P}_{\sigma_2}$ with Eq. (23) without the need for additional computational levels.

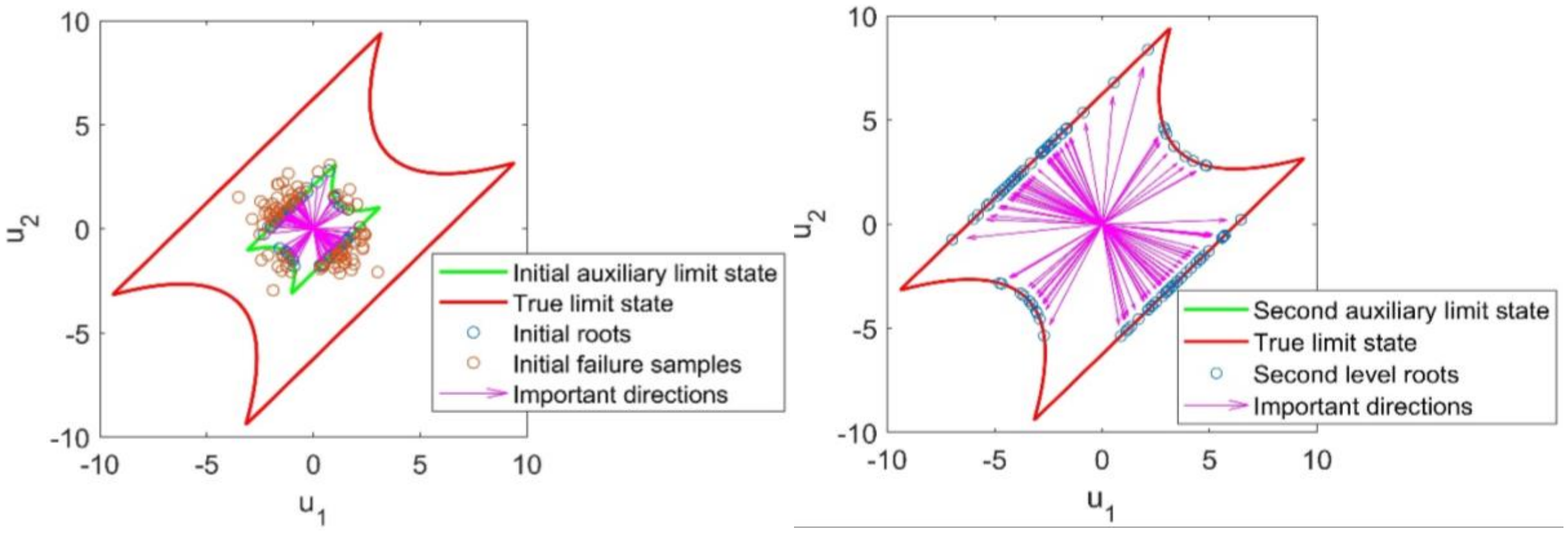

(**a**) First step (**b**) Second step

**Fig. 3**. Samples and important directions in the intermediate sampling steps of SDIS for example 4.1.

To investigate the performance of SDIS comprehensively, three different initial values of $\sigma_1$ are considered. Table 1 lists the mean number of model evaluations $E(N_{\text{tot}})$, mean estimate of the failure probability $E(\widehat{P}_f)$ and its CV $\delta(\widehat{P}_f)$ from the 300 independent runs obtained by three different methods. Additionally, the mean CV $E(\hat{\delta}_{\hat{P}_f})$ estimated with Eq. (30) using the intermediate samples from each simulation is also provided for comparison. In SIS, the independent M-H algorithm is utilized with $K = 4$. It is observed that the proposed SDIS outperforms SuS and SIS with respect to robustness and efficiency. In SuS and SIS, at least 5 sampling levels are required to obtain a convergent result. However, the proposed SDIS yields convergent failure probability with only 2 steps, and the important directions in SDIS generally capture all the significant regions of the failure domain. Moreover, we see that the estimate of the CV, $\hat{\delta}_{\hat{P}_f}$, evaluated with Eq. (30) is a good approximation of the true CV $\delta(\hat{P}_f)$ obtained with 300 independent runs.

**Table 1.** Reliability analysis results of Example 4.1 with reference value $P_f = 1.058\text{e}^{-5}$

| | SuS<br>CSM-H | SIS<br>IM-H ($K = 4$) | SDIS<br>$\sigma_1 = 3$ | <br>$\sigma_1 = 4$ | <br>$\sigma_1 = 5$ |
|---|---|---|---|---|---|
| $E(\widehat{P}_f)$ | 1.028e$^{-5}$ | 6.421e$^{-6}$ | 1.045e$^{-5}$ | 1.051e$^{-5}$ | 1.051e$^{-5}$ |
| $\delta(\widehat{P}_f)$ | 0.472 | 0.451 | 0.141 | 0.145 | 0.143 |
| $E(\widehat{\delta}_{\hat{P}_f})$ | - | - | 0.145 | 0.155 | 0.158 |
| $E(N_{\text{tot}})$ | 5358 | 5980 | 990.6 | 819.5 | 968.1 |

Additionally, we see from Table 1 that SDIS with initial value $\sigma_1 = 3$ works well, and a larger initial value of $\sigma_1$ ($\sigma_1 = 4$) can improve its efficiency further. However, for $\sigma_1 = 5$, another intermediate step may be (this situation occurs several times in the 500 independent runs) required to obtain a convergent estimation of the failure probability, which in turn undermines the performance of SDIS.

4.2 Two distinct failure regions

The second example is a series system defined by the following LSF [38] in standard normal space:

$$g(\boldsymbol{u}) = \min\left\{\begin{matrix} 3.2 + (u_1 + u_2)/\sqrt{2} \\ 2.5 + 0.1(u_1 - u_2)^2 - (u_1 + u_2)/\sqrt{2} \end{matrix}\right\} + 3. \tag{39}$$

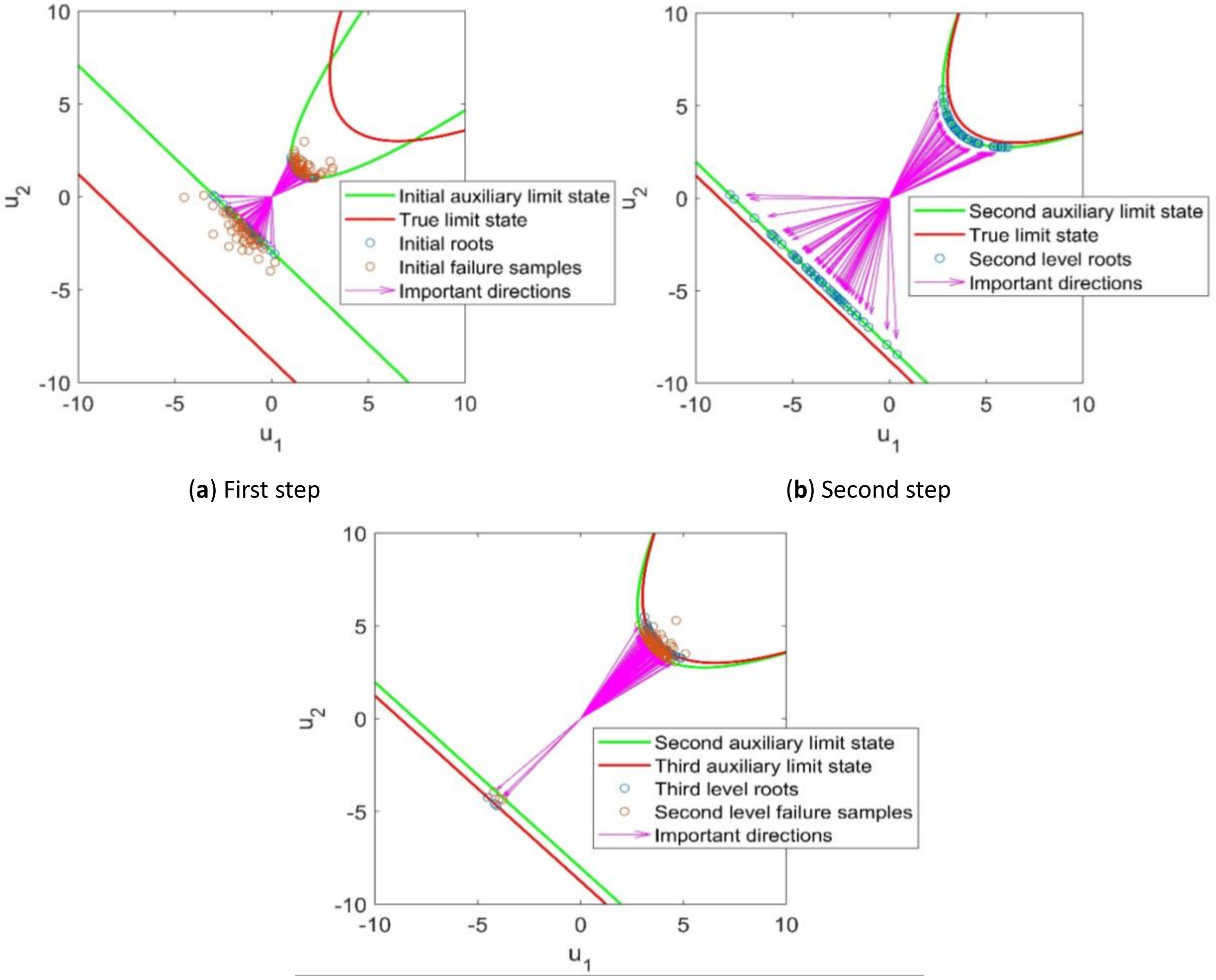

(**a**) First step (**b**) Second step

(**c**) Third step

**Fig. 4**. Samples and important directions in the intermediate sampling steps of SDIS for example 4.2.

The reference value of failure probability obtained by MCS with $10^{11}$ samples is $1.10 \times 10^{-8}$. In Fig. 4, a specific example run is used to illustrate the performance of SDIS in this example. Similar with Example 4.1, we firstly set $\sigma_1 = 3$, and generate samples following $\varphi_2(\boldsymbol{u})$ sequentially until 100 samples satisfying $G(\sigma_1 \boldsymbol{u}) \leq 0$ are obtained. By seeking the root of $G(\sigma_1 r \boldsymbol{a}) = 0$ along the directions defined by the failure samples, we obtain the weights corresponding to the 100 direction samples following $h_1(\boldsymbol{a})$, which allow us to estimate $P_{\sigma_2}$. In this example run, the optimal value of $\sigma_2$ obtained by Eq. (29) is 1.173, and the second auxiliary limit state surface $G(\sigma_2 \boldsymbol{u}) = 0$ is depicted in Fig. 4(b). Next, we perform the resample scheme described in Section 3.3 to draw 100 seeds, and run 100 Markov chains with length 5 independently. The last state of each Markov chain following $\tilde{h}_2(\boldsymbol{u})$ is depicted in Fig 4(c). By transforming these samples to polar coordinates, we obtain 100 samples of $\boldsymbol{a}$ following $h_2(\boldsymbol{a})$. The optimal value of $\sigma_3$ obtained with Eq. (29) is 1, and thus we have $\hat{P}_f = \hat{P}_{\sigma_3}$ in this example.

Table 2 compares the reliability analysis results obtained by various methods. For this challenging problem, SuS and SIS generally require 8 or 9 levels of samples to yield a convergent result. In contrast, SDIS gives an accurate estimation of failure probability with only three steps. As a result, SDIS requires much less model evaluations but provides much more robust result compared to SuS and SIS. Additionally, SDIS equipped with IM-H algorithm and CSM-H algorithm provides comparable results in this example. Since the failure probability in this example is quite small, increasing $\sigma_1$ will improve the efficiency of generating failure samples dramatically with MCS in the initial step of SDIS. Different from Example 4.1, even though additional intermediate steps may be required to yield a convergent result in this example, an obvious efficiency improvement can still be observed with larger $\sigma_1$, e.g., 4 and 5. Additionally, it is observed that $\delta_{\hat{P}_f}$ in Eq. (30) underestimates $\delta(\hat{P}_f)$ to some degree, especially for larger $\sigma_1$. This is attributed to the effect of the sample correlation, which is neglected in Eq. (32).

**Table 2.** Reliability analysis results of Example 4.2 with reference value $P_f = 1.10 \times 10^{-8}$

| | SuS CSM-H | SIS IM-H ($K = 2$) | SDIS ($l = 5, N_0 = 100$) IM-H ($K = 2$) | | | SDIS ($l = 5, N_0 = 100$) CSM-H | | |
|---|---|---|---|---|---|---|---|---|
| | | | $\sigma_1 = 3$ | $\sigma_1 = 4$ | $\sigma_1 = 5$ | $\sigma_1 = 3$ | $\sigma_1 = 4$ | $\sigma_1 = 5$ |
| $E(\hat{P}_f)$ | 1.061e$^{-8}$ | 8.824e$^{-9}$ | 1.110e$^{-8}$ | 1.082e$^{-8}$ | 1.095e$^{-8}$ | 1.089e$^{-8}$ | 1.066e$^{-8}$ | 1.101e$^{-8}$ |
| $\delta(\hat{P}_f)$ | 0.549 | 0.267 | 0.182 | 0.189 | 0.216 | 0.193 | 0.211 | 0.225 |
| $E(\hat{\delta}_{\hat{P}_f})$ | - | - | 0.178 | 0.182 | 0.183 | 0.179 | 0.182 | 0.183 |
| $E(N_{\text{tot}})$ | 7750 | 7970 | 3901.6 | 2407.3 | 2037.6 | 3924.5 | 2415.6 | 2029.8 |

4.3 Nonlinear oscillator

The third example investigates the performance of the SDIS method for nonlinear limit state problems with medium dimension. To this end, we consider a nonlinear oscillators model, with LSF defined as

$$g(\boldsymbol{x}) = 3r - \left|\frac{2F_1}{M\omega_0^2} sin\left(\frac{\omega_0 t_1}{2}\right)\right|, \tag{40}$$

where $\omega_0 = \sqrt{(c_1 + c_2)/M}$, and $\boldsymbol{x} = (M, c_1, c_2, r, F_1, t_1)^{\mathrm{T}}$ is the input vector. All the inputs are independently normally distributed random variables, and their distribution parameters are listed in Table 3. In this example, an isoprobabilistic transformation is used to transform the inputs into the standard normal space, and the MCS ($10^9$ samples) reference value of the failure probability $P_f$ is $6.43 \times 10^{-6}$.

**Table 3**. Distribution parameters of damped oscillator

| Variables | $M$ | $c_1$ | $c_2$ | $r$ | $F_1$ | $t_1$ |
|---|---|---|---|---|---|---|
| Mean | 1 | 1 | 0.1 | 0.5 | 0.3 | 1 |
| Standard deviation | 0.05 | 0.1 | 0.01 | 0.05 | 0.2 | 0.2 |

**Table 4.** Reliability analysis results of Example 4.3 with reference value $P_f = 6.43 \times 10^{-6}$

| | SuS CSM-H | SIS IM-H ($K = 1$) | SDIS ($l = 5, N_0 = 100$) IM-H ($K = 1$) | | | CSM-H | | |
|---|---|---|---|---|---|---|---|---|
| | | | $\sigma_1 = 3$ | $\sigma_1 = 4$ | $\sigma_1 = 5$ | $\sigma_1 = 3$ | $\sigma_1 = 4$ | $\sigma_1 = 5$ |
| $E(\widehat{P}_f)$ | 6.296e$^{-6}$ | 5.761e$^{-6}$ | 7.585e$^{-6}$ | 7.746e$^{-6}$ | 8.245e$^{-6}$ | 6.587e$^{-6}$ | 6.582e$^{-6}$ | 6.533e$^{-6}$ |
| $\delta(\widehat{P}_f)$ | 0.383 | 0.195 | 0.253 | 0.282 | 0.301 | 0.291 | 0.336 | 0.331 |
| $E(\widehat{\delta}_{\widehat{P}_f})$ | - | - | 0.208 | 0.230 | 0.234 | 0.211 | 0.232 | 0.236 |
| $E(N_{\text{tot}})$ | 5934 | 6000 | 2403.4 | 2411.5 | 2384.3 | 2423.9 | 2486.4 | 2396.9 |

Table 4 lists the reliability analysis results obtained with different methods. This model only contains a single failure domain, thus SIS and SDIS coupled with IM-H algorithm with $K = 1$ yield the lower CV, but these results are slightly biased. SDIS combined with the CSM-H algorithm provides accurate and robust estimation of failure probability with nearly half the model evaluations compared to SuS and SIS. In this example, the improvement of computational efficiency is marginal by increasing $\sigma_1$, but the CV $\delta(\widehat{P}_f)$ increases significantly since more intermediate steps are required in SDIS to yield convergent estimate of the failure probability. In addition, one can see that $\hat{\delta}_{\hat{P}_f}$ again underestimates $\delta(\widehat{P}_f)$.

4.4 Linear high dimensional problem

In the fourth example we test the performance of SDIS in a high-dimensional problem. To this end, we consider the following LSF [46]

$$g(\boldsymbol{u}) = \beta - \frac{1}{\sqrt{n}}\sum_{i=1}^{n} u_i, \tag{41}$$

where all the inputs are independent standard normal distributed random variables. In this example, we set $\beta$ as 4, and thus the true failure probability is given by $P_f = \Phi(-\beta) = 3.167 \times 10^{-5}$.

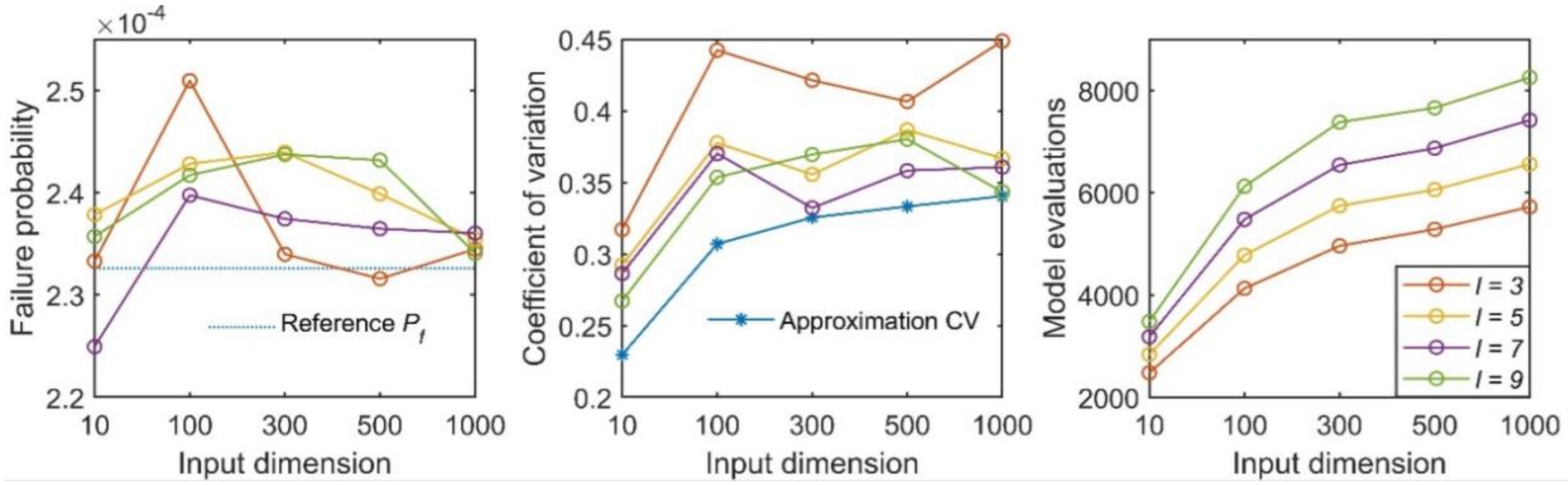

(a) (b) (c)

**Fig. 5.1** SDIS with $N_0$ =100 important directions per level

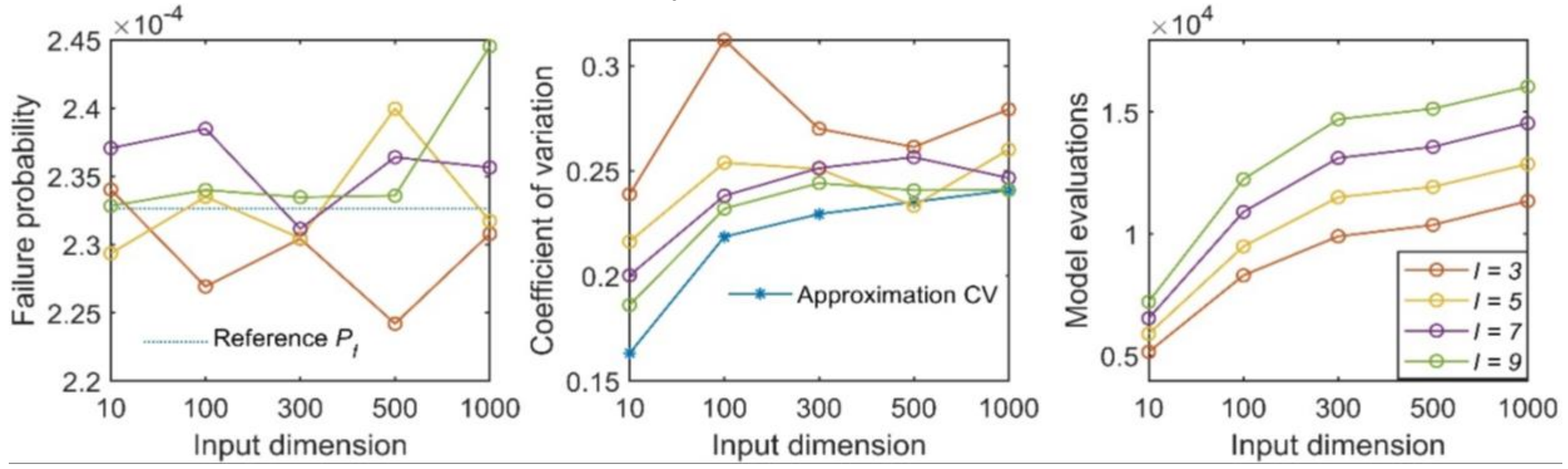

(**a**) (**b**) (**c**)

**Fig. 5.2** SDIS with $N_0$ =200 important directions per level

**Fig. 5** Performance of SDIS with different Markov chain length $l$ of Example 4.4; (**a**) Failure probability against input dimension; (**b**) CV of failure probability against input dimension; (**c**) Computational costs of SDIS against input dimension.

In this example, the performance of SDIS is tested for varying number of random variable ($n = 10{,}100{,}300{,}500$ and 1000). Meanwhile, different number of important directions per level ($N_0 = 100$ and 200) in SDIS is considered, and the effect of the burn-in stage is investigated by setting the length $l$ of each Markov chain as 3, 5, 7 and 9, respectively. As depicted in Fig. 5, the failure probability estimates obtained with SDIS are close to the true probability for all different considered settings. The computational cost of SDIS increases gradually as a function of the input dimension. Indeed, with increase of the input dimension, more and more intermediate levels are generally required to yield a convergent estimation of the failure probability. As a result, the CV of SDIS estimator increases with increase of the input dimension (cf. Eq. (32)). However, in very high-dimensional problems ($n > 100$), the rate of increase of the total number of levels in SDIS decreases, and the CV of SDIS estimator tends to stabilize. Moreover, it is observed that the CV of SDIS estimator in Eq. (30) can describe the trend of the true CV, but it generally underestimates the true one. The difference between the approximated CV and the true one can be reduced by increasing the length $l$ of Markov chain, and we see that $l = 5$ already provides relatively good result. Additionally, we see that the CV of SDIS estimator decreases with increase of $N_0$ from 100 to 200, at the expense of an increase of the computational cost.

Table 5 compares the detailed reliability analysis results of the different methods. It is observed that all these methods are applicable to problems with various dimensions. Specifically, SDIS outperforms other methods for problems with medium dimension. However, SuS yields smaller CV than SIS and SDIS at lower computational cost with the same MCMC sampling algorithm for $n > 100$, which suggests that SuS is superior to the other two methods in very high dimensional problems.

**Table 5.** Reliability analysis results of Example 4.4 ($2.326e^{-4}$)

| $n = 10$ | SuS CSM-H | SIS CSM-H | SDIS ($\sigma_1 = 3, N_0 = 100$) CSM-H | | | |
|---|---|---|---|---|---|---|
| | | | $l = 3$ | $l = 5$ | $l = 7$ | $l = 9$ |
| $E(\widehat{P}_f)$ | $2.404e^{-4}$ | $2.358e^{-4}$ | $2.333e^{-4}$ | $2.379e^{-4}$ | $2.249e^{-4}$ | $2.357e^{-4}$ |
| $\delta(\widehat{P}_f)$ | 0.292 | 0.410 | 0.317 | 0.292 | 0.286 | 0.267 |
| $E(\widehat{\delta}_{\widehat{P}_f})$ | - | - | 0.230 | 0.230 | 0.230 | 0.230 |
| $E(N_{\text{tot}})$ | 4458.6 | 4966.6 | 2479.6 | 2838.2 | 3184.1 | 3482.8 |
| $n = 100$ | SuS CSM-H | SIS CSM-H | SDIS ($\sigma_1 = 3, N_0 = 100$) CSM-H | | | |
| | | | $l = 3$ | $l = 5$ | $l = 7$ | $l = 9$ |
| $E(\widehat{P}_f)$ | $2.431e^{-4}$ | $2.275e^{-4}$ | $2.509e^{-4}$ | $2.428e^{-4}$ | $2.398e^{-4}$ | $2.417e^{-4}$ |
| $\delta(\widehat{P}_f)$ | 0.293 | 0.390 | 0.442 | 0.377 | 0.370 | 0.353 |
| $E(\widehat{\delta}_{\widehat{P}_f})$ | - | - | 0.307 | 0.307 | 0.307 | 0.307 |
| $E(N_{\text{tot}})$ | 4455.8 | 4983.3 | 4129.1 | 4780.8 | 5478.9 | 6124.9 |
| $n = 1000$ | SuS CSM-H | SIS CSM-H | SDIS ($\sigma_1 = 3, N_0 = 100$) CSM-H | | | |
| | | | $l = 3$ | $l = 5$ | $l = 7$ | $l = 9$ |
| $E(\widehat{P}_f)$ | $2.375e^{-4}$ | $2.308e^{-4}$ | $2.344e^{-4}$ | $2.355e^{-4}$ | $2.361e^{-4}$ | $2.341e^{-4}$ |
| $\delta(\widehat{P}_f)$ | 0.312 | 0.406 | 0.448 | 0.367 | 0.361 | 0.343 |
| $E(\widehat{\delta}_{\widehat{P}_f})$ | - | - | 0.340 | 0.340 | 0.340 | 0.340 |
| $E(N_{\text{tot}})$ | 4466.9 | 4986.7 | 5723.2 | 6558.7 | 7418.8 | 8253.4 |

4.5 High dimensional series system problem

The fifth example tests the performance of the SDIS method for a high-dimensional series system problem of two linear LSFs in opposite sides of the origin in a standard normal input space [23]:

$$g(\boldsymbol{u}) = \min \begin{Bmatrix} \beta - \frac{1}{\sqrt{n}} \sum_{i=1}^{n} u_i \\ \beta + \frac{1}{\sqrt{n}} \sum_{i=1}^{n} u_i \end{Bmatrix}. \tag{42}$$

Again, here we set $\beta$ as 4, and thus the true failure probability for this problem is given by $P_f = 2\Phi(-\beta) = 6.334e^{-5}$.

Similar with Example 4, here we test the performance of SDIS with different settings, and the results are presented in Fig. 6. It is shown that SDIS yields nearly unbiased estimation of failure probability for all cases. Again, we see that the CV of SDIS estimator increases fast with increase of the input dimension and tends to stabilize as $n > 100$.

The comparisons of the reliability analysis results are reported in Table 6. Although the failure domain in this example has two unconnected failure regions, we see that all considered methods give accurate results at various dimensions. Again, SDIS is superior than other methods for problem with medium dimension, and SuS provides the smallest CV at lowest computational cost for very high-dimensional cases.

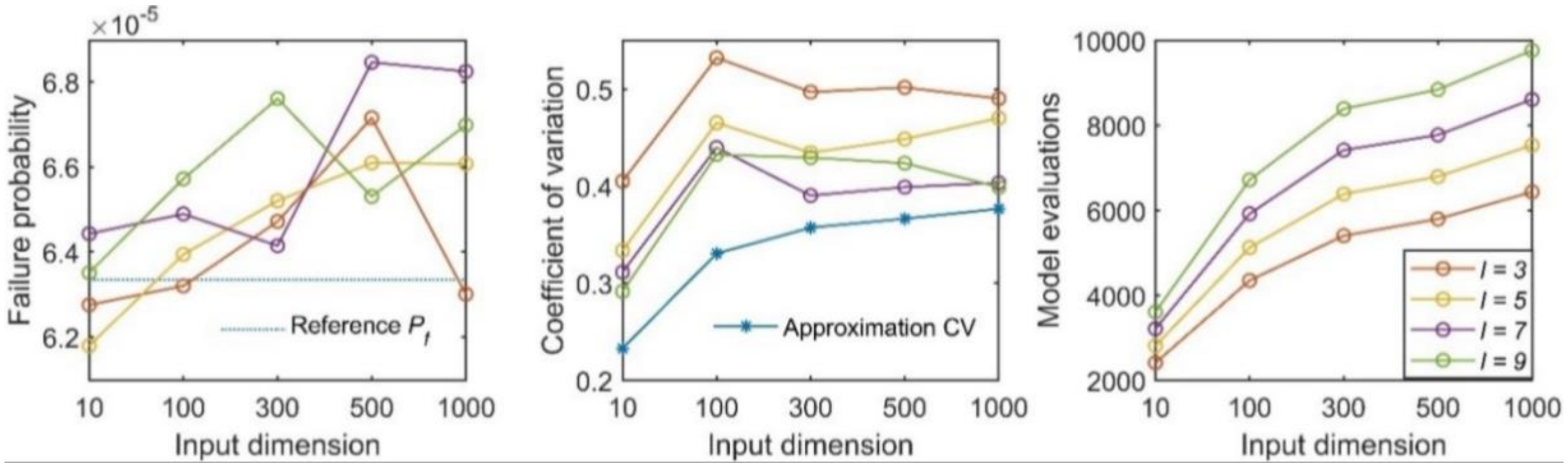

(**a**) (**b**) (**c**)

**Fig. 6.1**. SDIS with $N_0$ =100 important directions per level

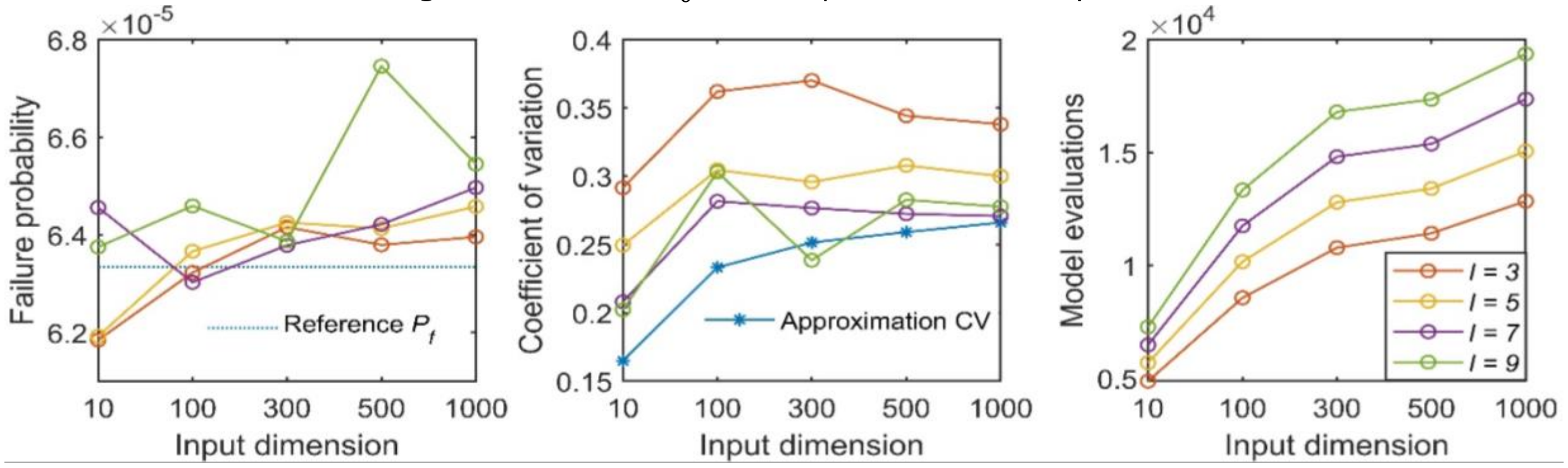

(**a**) (**b**) (**c**)

**Fig. 6.2**. SDIS with $N_0$ =200 important directions per level

**Fig. 6** Performance of SDIS with different Markov chain length $l$ of Example 4.5; (**a**) Failure probability against input dimension; (**b**) CV of failure probability against input dimension; (**c**) Computational costs of SDIS against input dimension.

**Table 6.** Reliability analysis results of Example 4.5 (Reference $P_f = 6.334\mathrm{e}^{-5}$)

| $n = 10$ | SuS CSM-H | SIS CSM-H | SDIS ($\sigma_1 = 3, N_0 = 100$) CSM-H | | | |
|---|---|---|---|---|---|---|
| | | | $l = 3$ | $l = 5$ | $l = 7$ | $l = 9$ |
| $E(\widehat{P}_f)$ | 6.409e$^{-5}$ | 6.248e$^{-5}$ | 6.275e$^{-5}$ | 6.179e$^{-5}$ | 6.443e$^{-5}$ | 6.351e$^{-5}$ |
| $\delta(\widehat{P}_f)$ | 0.317 | 0.515 | 0.405 | 0.334 | 0.311 | 0.291 |
| $E(\widehat{\delta}_{\widehat{P}_f})$ | - | - | 0.233 | 0.233 | 0.233 | 0.233 |
| $E(N_{\mathrm{tot}})$ | 4827.2 | 5010 | 2415.2 | 2815.7 | 3214.8 | 3609.8 |

| $n = 100$ | SuS CSM-H | SIS CSM-H | SDIS ($\sigma_1 = 3, N_0 = 100$) CSM-H | | | |
|---|---|---|---|---|---|---|
| | | | $l = 3$ | $l = 5$ | $l = 7$ | $l = 9$ |
| $E(\widehat{P}_f)$ | 6.576e$^{-5}$ | 6.237e$^{-5}$ | 6.320e$^{-5}$ | 6.394e$^{-5}$ | 6.489e$^{-5}$ | 6.572e$^{-5}$ |
| $\delta(\widehat{P}_f)$ | 0.353 | 0.504 | 0.532 | 0.465 | 0.439 | 0.432 |
| $E(\widehat{\delta}_{\widehat{P}_f})$ | - | - | 0.331 | 0.331 | 0.331 | 0.331 |
| $E(N_{\mathrm{tot}})$ | 4899.6 | 5010 | 4352.4 | 5126.1 | 5920.4 | 6720.6 |

| $n = 1000$ | SuS CSM-H | SIS CSM-H | SDIS ($\sigma_1 = 3, N_0 = 100$) CSM-H | | | |
|---|---|---|---|---|---|---|
| | | | $l = 3$ | $l = 5$ | $l = 7$ | $l = 9$ |
| $E(\widehat{P}_f)$ | $6.335e^{-5}$ | $6.506e^{-5}$ | $6.300e^{-5}$ | $6.607e^{-5}$ | $6.824e^{-5}$ | $6.698e^{-5}$ |
| $\delta(\widehat{P}_f)$ | 0.347 | 0.538 | 0.490 | 0.470 | 0.404 | 0.399 |
| $E(\widehat{\delta}_{\widehat{P}_f})$ | - | - | 0.376 | 0.376 | 0.376 | 0.376 |
| $E(N_{\text{tot}})$ | 4935.2 | 5020 | 6436.7 | 7535.8 | 8613.5 | 9766.8 |

## 5. Conclusions and future works

This paper proposed a sequential directional importance sampling (SDIS) method for rare event estimation. SDIS expresses the probability of failure in terms of the probabilities of a sequence of auxiliary reliability problems, defined by magnifying the input variability. The first probability is estimated with Monte Carlo simulation in Cartesian coordinates, while the subsequent ones are computed in polar coordinates through a directional importance sampling technique. For the latter, each directional importance sampling density is chosen as the optimal sampling density of the previous auxiliary failure probability in the sequence. Samples from the directional importance densities are generated sequentially through a resample-move scheme, for which a novel resampling strategy and two tailored Markov chain Monte Carlo algorithms are introduced. Finally, an adaptive approach for selecting the intermediate auxiliary problems is presented and the variance of the resulting SDIS estimator is analyzed.

Five benchmark examples are used to assess the performance of the proposed method. The results show that SDIS provides robust and accurate reliability analysis results for problems with nonlinear limit state functions, multiple failure domains, small failure probabilities and various dimensions of the input space. SDIS is shown to outperform other sequential sampling approaches, namely subset simulation (SuS) and sequential importance sampling (SIS), in low- to moderately high-dimensional problems . Even though the performance of SDIS is shown to deteriorate with increase of the problem dimension, it is still applicable for very high dimensional problems. However, our studies suggest that SuS remains the method of choice at dimensions > 100 .

The initial magnification factor $\sigma_1$, the burn-in period $l$ of the Markov chains and the number of important directions $N_0$ per level have strong influence on the quality of the estimates obtained by the proposed SDIS method. The studies in this paper showed that the optimal choice of $\sigma_1$ is problem dependent, with $\sigma_1 = 3$ giving acceptable results. SDIS with larger $l$ gives smaller coefficient of variation of the failure probability estimate at the expense of an increase in computational cost; we observe that $l = 5$ provides a good compromise. To further improve the performance of SDIS, future work lies in determining the parameter $\sigma_1$ adaptively and exploring more advanced line search algorithms. Another possible future direction is the adaptive choice of the number of sampling directions $N_0$ used to estimate the intermediate probabilities.

## Acknowledgements

This work was supported by the National Natural Science Foundation of China (Grant No. NSFC 11902254), National Science and Technology Major Project (Grant No. 2017-IV-0009-0046).